\documentstyle[twoside,fleqn,espcrc2,epsfig ]{article}

\def\lsim{\mathrel{\rlap{\lower4pt\hbox{\hskip1pt$\sim$}}
    \raise1pt\hbox{$<$}}}         %less than or approx. symbol
\def\gsim{\mathrel{\rlap{\lower4pt\hbox{\hskip1pt$\sim$}}
    \raise1pt\hbox{$>$}}}         %greater than or approx. symbol

\def\frac#1#2{{{#1}\over {#2}}}

\def\NP{{\it Nucl.~Phys.~}}

\def\vol#1{{\bf #1}}\def\vyp#1#2#3{\vol{#1} (#2) #3}

\newcommand{\AmS}{{\protect\the\textfont2
  A\kern-.1667em\lower.5ex\hbox{M}\kern-.125emS}}

\begin{document}
\begin{titlepage}
\setcounter{page}{0}
\topmargin 2mm
\begin{flushright}
{\tt hep-ph/9808462}\\
{CERN-TH/98-285}\\
{DFTT 55/98}\\
\end{flushright}
%\vskip 12pt
\begin{center}
{\bf ARE PARTON DISTRIBUTIONS POSITIVE?} 
\vskip 12pt
{Stefano Forte\footnote[1]{Address after September 1, 1998: INFN, Sezione di Roma
Tre, via della Vasca Navale 84, I-00146 Rome, Italy}}\\
\vskip 6pt
{\em INFN, Sezione di Torino, via P.~Giuria 1, I-10125 Torino, Italy }\\ 
\vskip 10pt
{Guido Altarelli}\\
\vskip 6pt
{\em Theory Division, CERN CH-1211 Geneva 23,
Switzerland} \\ 
{\em Dipartimento di Fisica ``E.~Amaldi'', 
Universit\`a di Roma Tre, I-00146 Rome, Italy}\\
\vskip 5pt
{and}
\vskip 5pt
 
{Giovanni Ridolfi}\\
{\em INFN,
Sezione di Genova, via Dodecaneso 33, I-16146 Genova, Italy}\\
\vskip 30pt
\end{center}
{\narrower\narrower
\centerline{\bf Abstract}
\medskip\noindent
We show that the naive positivity conditions on polarized parton
distributions which follow from their probabilistic interpretation in 
the naive parton model are reproduced in
perturbative QCD  at the leading log level if the quark and
gluon distribution are defined in terms of physical processes. We show how
these conditions are modified at the next-to-leading level, and  discuss
their phenomenological implications, 
in particular in view of the
determination of the polarized gluon distribution.}

\smallskip
\vskip 60pt
\begin{center}
{Invited talk given at\\\smallskip {\em QCD98}\\ Montpellier, July 1996}\\
\smallskip
{\em to be published in the proceedings}\\
\end{center}
\bigskip
\vfill
\begin{flushleft}
{CERN-TH/98-285}\\
{August 1998}
\end{flushleft}
\end{titlepage}

%RRRRRRRRRRRRRRRRRRRRRRRRRRRRRRRRRRRRRRRRRRRRRRRRRRRRRRRRRRRRRRRRRRRRRRRRRRR%

\title{Are Parton Distributions Positive?
                                      \thanks{Presented by SF}}

\author{Stefano~Forte\address{INFN,
Sezione di Torino, via P.~Giuria 1, I-10125 Torino,
Italy}\thanks{Address after September 1, 1998: INFN, Sezione di Roma
Tre, via della Vasca Navale 84, I-00146 Rome, Italy},
Guido~Altarelli\address{Theory Division, CERN, CH-1211 Geneva 23,
Switzerland \\$\>$ 
Dipartimento di Fisica ``E.~Amaldi'', 
Universit\`a di Roma Tre, I-00146 Rome, Italy}, 
Giovanni Ridolfi\address{INFN,
Sezione di Genova, via Dodecaneso 33, I-16146 Genova, Italy}}

\begin{abstract}

We show that the naive positivity conditions on polarized parton
distributions which follow from their probabilistic interpretation in 
the naive parton model are reproduced in
perturbative QCD  at the leading log level if the quark and
gluon distribution are defined in terms of physical processes. We show how
these conditions are modified at the next-to-leading level, and  discuss
their phenomenological implications, 
in particular in view of the
determination of the polarized gluon distribution.

\end{abstract}

\maketitle

In the naive parton model, parton distributions are viewed as
probabilities for a parton with given momentum fraction to be found in
its parent hadron, and are thus positive
semi--definite.
When extended to the polarized case, this simple picture 
leads to nontrivial constraints:  since the polarized and
unpolarized distributions $\Delta f$ and $f$ are sums and differences 
of helicity distributions, 
\begin{equation}
f=f^\uparrow+f^\downarrow,\quad
\Delta =f^\uparrow-f^\downarrow,
\label{pdfdef}\end{equation}
it follows that
\begin{equation}
|\Delta f(x,Q^2)|\le f(x,Q^2).
\label{naivebound}
\end{equation}

A minute's reflection shows that
these relations cannot be true to all orders in perturbative QCD in a
generic factorization scheme, since by choosing a scheme
we are free to change $f$ and $\Delta f$ independently of each other
in an arbitrary way, and thus the bound eq.(\ref{naivebound}) can be
violated at will. This raises the question whether in perturbative QCD
the bounds  
eq.(\ref{naivebound}) are valid at all, and more in
general whether the helicity distributions of eq.(\ref{pdfdef}) 
are necessarily positive semi--definite.

Let us consider first the simple case of the quark and antiquark
distributions. These  can be easily 
defined in terms of a physical
process: positivity of the parton
distribution can then be expressed in terms of the 
positivity of a physical cross section.
Specifically, we can express the unpolarized and polarized
deep-inelastic structure functions $F_1(x,Q^2)$ and $g_1(x,Q^2)$ in terms
of quark, antiquark and gluon distributions as
\begin{eqnarray}
&&F_1(x,Q^2)={{1}\over{2}}
\sum_{i=1}^{n_f} e^2_i C^d_i\otimes \left(q_i+\bar q_i\right)\nonumber
\\&&\qquad\qquad+ 
2n_f \langle e^2\rangle C^d_g \otimes 
g\\
&&g_1(x,Q^2)={{1}\over{2}}
\sum_{i=1}^{n_f} e^2_i \Delta C^d_i\otimes \left(\Delta q_i+\Delta
\bar q_i\right)\nonumber\\&&\qquad\qquad+
2n_f\langle e^2\rangle\Delta C^d_g \otimes \Delta g\nonumber,
\label{sfdef}
\end{eqnarray}
where $C_i$ and $\Delta C_i$ are perturbatively computable coefficient
functions.

Since the structure functions are in turn related to the 
asymmetry $A_1$ for deep-inelastic
scattering of transversely
polarized virtual photons on a longitudinally polarized nucleon 
through
\begin{equation}A_1\equiv
{\sigma_{1/2}-\sigma_{3/2}\over\sigma_{1/2}+\sigma_{3/2}}={g_1(x,Q^2)\over
F_1(x,Q^2)},\label{aonedef}
\end{equation}
it follows  that
$g_1$ is bounded by $F_1$:
\begin{equation}|g_1(x,Q^2)|\leq
F_1(x,Q^2)\label{sfbound}\end{equation}

But at the leading log (LL) level all the quark coefficient functions in
eq.(\ref{sfdef}) are equal to unity while all the gluon coefficient
functions vanish: 
\begin{eqnarray}
&&C(x,\alpha_s)=\sum_{k=0}^\infty 
\left(\alpha_s \over 2\pi\right)^k
C^{(k)}(x);\label{cfexp}\\
&&C^{d,\,(0)}_{i}(x)=\Delta
C^{d,\,(0)}_{i}(x)=\delta(1-x),\label{locfqq}\\
&&C^{d,\,(0)}_{g}(x)=\Delta C^{d,\,(0)}_{g}(x)=0.\label{locfqg} 
\end{eqnarray}
It follows that  the bound eq.(\ref{sfbound}) immediately
translates into the positivity condition
\begin{equation}|\Delta q_i(x,Q^2)|\leq
q_i(x,Q^2),\label{pdfbound}\end{equation}
where the bound on each quark flavor separately is obtained by
imposing that eq.(\ref{sfbound}) be true for any choice of target,
i.e. for any value of the quark charges $e^2_i$ in eq.(\ref{sfdef}).

Beyond the LL level, the partonic positivity condition
eq.(\ref{pdfbound}) will only be satisfied in a parton
scheme~\cite{partsch}, 
where
all coefficient functions retain their leading-order (LO) form
eqs.(\ref{locfqq}-\ref{locfqg}). 
Such a scheme choice
is however not necessarily advisable, since it preserves some
arbitrarily selected partonic properties at the
expense of other properties.
In a generic scheme, instead, the partonic positivity condition
will only hold at the LL level. Beyond LO,
eq.(\ref{pdfbound}) will in general be violated, and it will be
replaced by a generalized condition,
obtained substituting the full expression of $g_1$ and
$F_2$ eq.(\ref{sfdef}) in the bound which relates them
eq.(\ref{sfbound}). Accordingly, in a generic scheme the helicity
parton distributions $f^{\uparrow\downarrow}$ eq.(\ref{pdfdef}) 
are in general not
positive semi-definite beyond  LO.

Notice that this violation of positivity of parton distributions
beyond LO is peculiar of the  polarized
sector. Indeed, in the unpolarized case the contributions $C^{(i)}$ to the
coefficient functions beyond the first order
need not be positive semi-definite --- for example, all Mellin moments
\begin{equation}
C(N,Q^2)\equiv\int_0^1\!dx\, x^{N-1} C(x,Q^2)\label{meldef}\end{equation}
of $C^{d,\,(1)}_q(x)$ with $N>1$  are negative. However, this
is merely a subleading correction to the LO unpolarized
coefficient function, thus within the region of
validity of perturbation theory the full coefficient function
$C(x,Q^2)$ remains positive semi-definite, and so does 
the unpolarized quark distribution as a consequence of the
positiveness of the structure function $F_1$. 
On the other hand, whenever the bound eq.(\ref{sfbound}) is violated,
either 
$f^\uparrow< 0$ or $f^\downarrow< 0$ : this may for
instance happen   if $\Delta
C_i(x,Q^2) > C_i(x,Q^2)$
while the gluon contributions in eq.(\ref{pdfdef})
are negligible.

In order  to extend the above discussion
to the case of the gluon distribution, we need a pair
of polarized and unpolarized processes which at LO define the
gluon, i.e. such that their LO coefficient
functions are given by
\begin{eqnarray}
&&C^{h,\, (0)}_{i}(x)=\Delta
C^{h,\, (0)}_{i}(x)=0
\label{locfgg}\\
&&C^{h,\, (0)}_{g}(x)=\Delta C^{h,\,
(0)}_{g}(x)=\delta(1-x).\label{locfgq}
\end{eqnarray}
An adequate choice~\cite{pos} is the 
inclusive production of a scalar particle, such
as the Higgs, in gluon--proton collisions (fig.~1). 
\begin{figure}[t]
\vbox{
\epsfig{figure=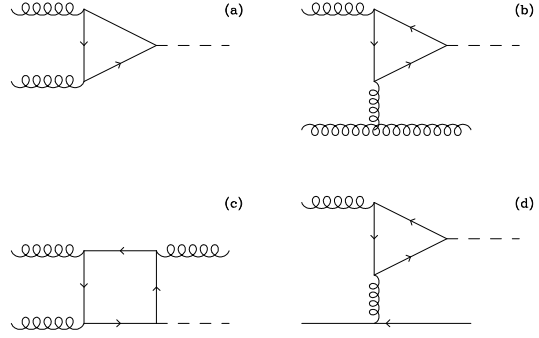,width=7.truecm}}
\vskip-.6truecm
\caption{\footnotesize 
 Diagrams for the processes (a)  $g+g\to H$,
(b) $g+q\to H+q$, (c, d) $g+g\to H+g$.}\vskip-.7truecm
\end{figure}
The unpolarized
and polarized cross--sections for this process can be written as
convolutions of partonic cross sections  with parton distributions in
the proton:
\begin{eqnarray}
&&\sigma[gp\to H+X](x,m_h^2)=A\left[C^h_g\otimes
g\right.\nonumber\\&&\qquad\qquad\left.+C^h_s\otimes\Sigma\right]\\
&&\Delta\sigma[gp\to H+X](x,m_h^2)=A\left[\Delta C^h_g\otimes \Delta
g\right.\nonumber\\
&&\qquad\qquad\left.+
\Delta C^h_s\otimes
\Delta \Sigma\right]\nonumber\label{hsfdef}\end{eqnarray}
where $A$ is a dimensionful coefficient, 
$x\equiv{m_h^2\over s}$ and the singlet quark distribution is defined
as $\Sigma(x,Q^2) =\sum_{i=1}^{n_f} \left[q_i(x,Q^2)+\bar
 q_i(x,Q^2)\right]$ 
(and likewise for the polarized distribution $\Delta \Sigma$).
Now at LO (fig.~1a)
momentum conservation implies that the partonic cross section is
$\sigma(gg\to H)=C\delta(x s-M^2_H)$, 
while helicity
conservation implies that the helicity of the incoming gluon and the 
gluons in the proton must be the same (in a collinear frame).
This means that the LO coefficient
functions have the form eq.(\ref{locfgg}-\ref{locfgq}), and then
the requirement
\begin{equation}
|\Delta\sigma[gp\to H+X]|\leq \sigma[gp\to H+X]
\label{xsgbound}
\end{equation}
implies that the polarized and unpolarized 
gluon distributions satisfy the bound eq.(\ref{naivebound}).
 Beyond
LO more partons can be emitted in the final state (fig.~1b-d),
the coefficient functions are generic, and the LO bound is
again replaced by a generalized condition,
found using the explicit expression of the cross
section eq.(\ref{hsfdef}) in the cross-section positivity bound 
eq.(\ref{xsgbound}).

At next-to-leading order (NLO) 
the pair of naive bounds of the form eq.(\ref{naivebound})
satisfied by the quark singlet and gluon distribution are thus
replaced by a pair of bounds which mix these two distributions through
the coefficient functions of the processes that define them at LO. 
The bounds simplify by taking moments [as in eq.(\ref{meldef})],
which turn the
convolutions
in eqs.(\ref{sfdef},\ref{hsfdef}) into ordinary products: for instance, 
the NLO bound on the gluon becomes
\begin{equation}
{{\left| \Delta g\left(1+ {\alpha_s\over 2\pi} \Delta 
C^{h,\,(1)}_g\right)+{\alpha_s\over 2\pi} \Delta 
C^{h,\,(1)}_s
\Delta \Sigma\right| \over
  g\left(1+ {\alpha_s\over 2\pi}  
C^{h,\,(1)}_g\right)+{\alpha_s\over 2\pi} 
C^{h,\,(1)}_s
\Sigma}
\leq 1},
\label{nlogbound}
\end{equation}
where the various parton distributions and coefficient functions
depend on the Mellin variable $N$ and the scale $Q^2$.

Given the unpolarized parton distributions, which are in general much
better known than the polarized ones, the
bounds thus take the form of allowed areas in the $\Delta
\Sigma(N)$,$\Delta g(N)$ plane. These can be determined to NLO
with the above choice of defining processes, since the DIS coefficient
functions are well known~\cite{discf}, while the
unpolarized~\cite{higgs}
and polarized~\cite{pos} coefficient functions for the process of
fig.~1 have been determined recently. 
 The bounds (using CTEQ4LQ~\cite{cteq}  
unpolarized distributions)
are shown in fig.~2
for selected values of $N$. 
\begin{figure}[t]
\vbox{
\hbox{\hskip.8truecm\epsfig{figure=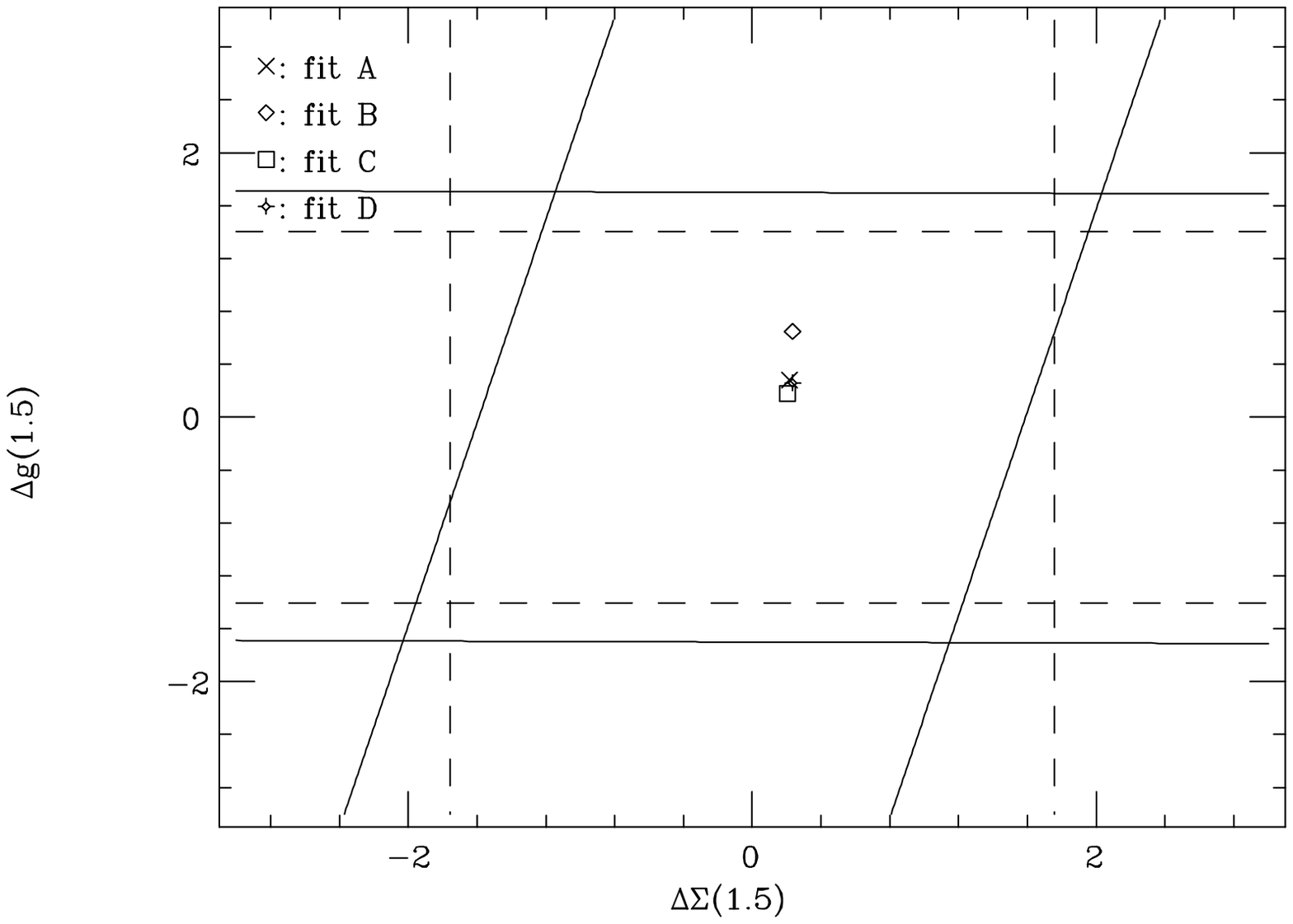,width=5.truecm}}
\vskip.2truecm\hbox{\hskip.8truecm\epsfig{figure=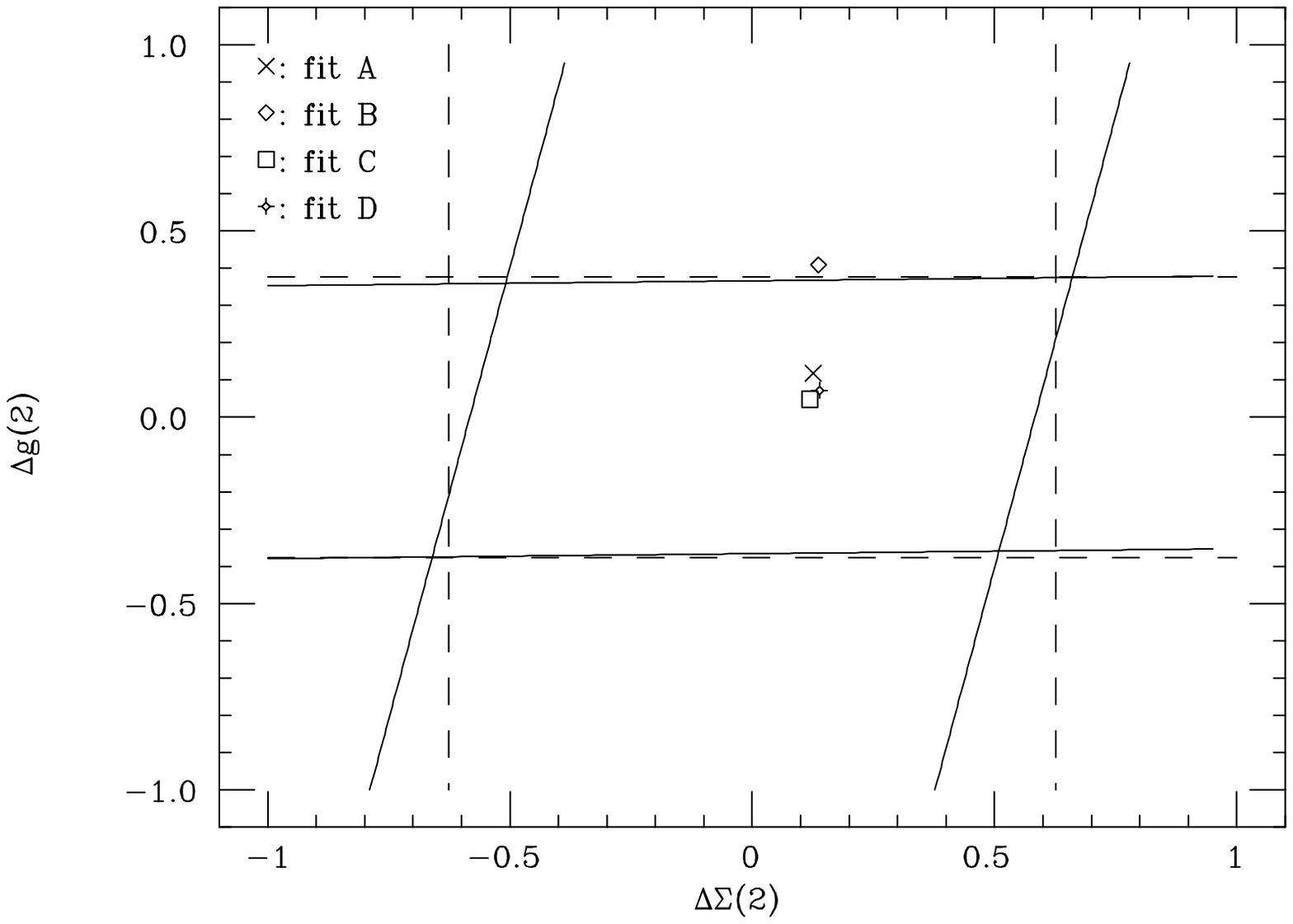,width=5.truecm}}
\vskip.2truecm\hbox{\hskip.8truecm\epsfig{figure=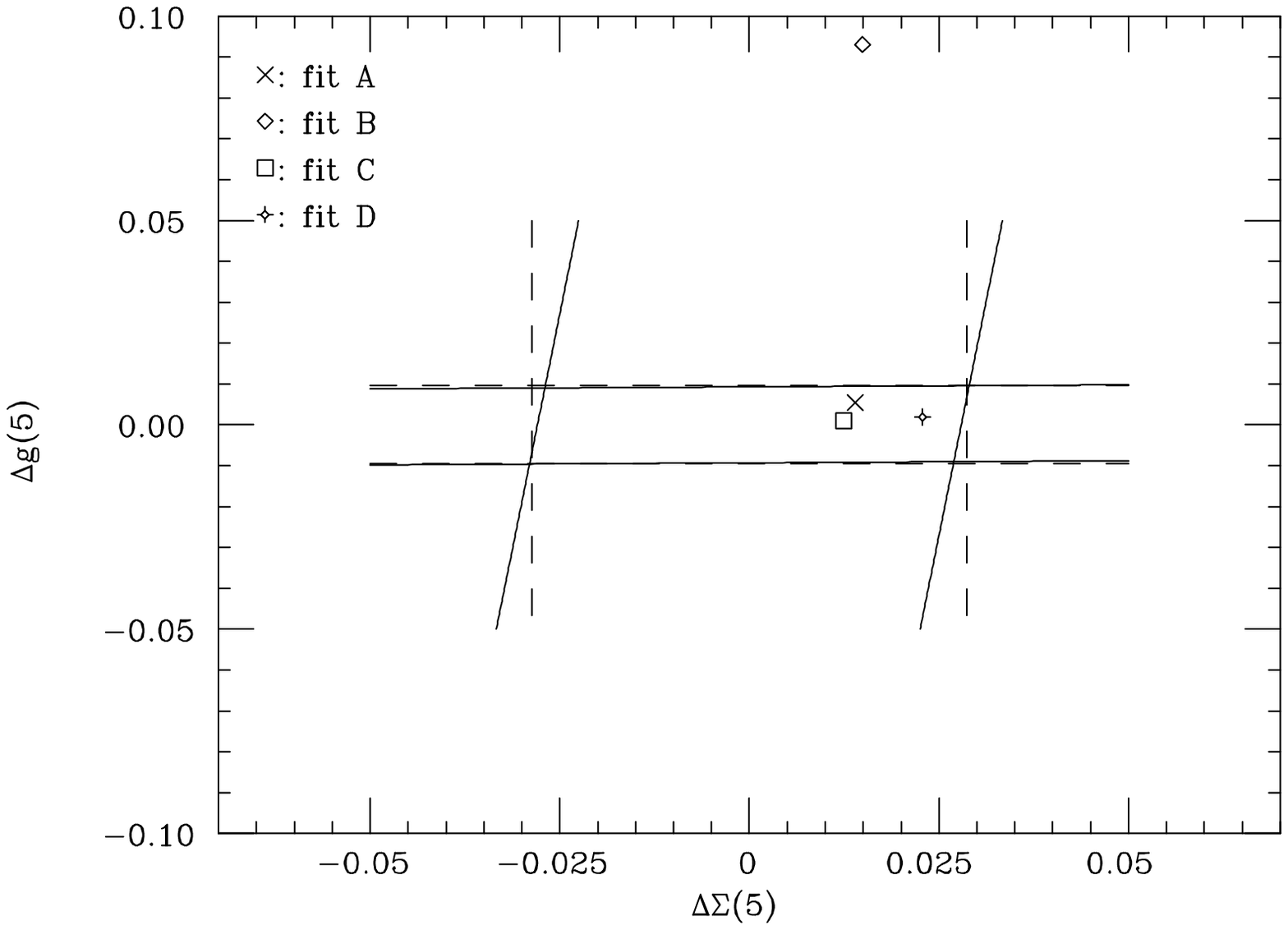,width=5.truecm}}}
\vskip-.6truecm
\caption{\footnotesize 
The LO (dashed lines) and NLO (solid lines) positivity
bounds on $\Delta \Sigma(N)$ and $\Delta g(N)$ for $Q^2=1$~GeV$^2$ and
$N=1.5,\,2,\,5$.
The values of $\Delta \Sigma$ and $\Delta g$ corresponding to the  NLO
fits  to $g_1$ data
of ref.~[6] are also shown.}\vskip-.9truecm
\end{figure}
The bounds are only significant for
$N>1$, because at  $N=1$ the unpolarized parton
distributions in the denominator of
eq.(\ref{nlogbound}) diverge. Because the unpolarized coefficient
functions in the denominator of
eq.(\ref{nlogbound}) also diverge as $N\to 1$, the
NLO bound on $\Delta g$ is much less restrictive than the naive LO one if  
$N$ is close to one. However, when $N$ grows the NLO bound
becomes somewhat more restrictive than the LO one, and eventually for very
large $N$ the difference between LO and NLO is negligible.

We may further use the bounds on the moments of the polarized gluon
distribution to construct a maximal gluon distribution, i.e. $\Delta
g_{\rm max} (x,Q^2)$  such that $|\Delta
g (x,Q^2)|\leq \Delta
g_{\rm max} (x,Q^2)$, by fixing
the polarized quark distribution and performing numerically the Mellin
inversion of the bound eq.(\ref{nlogbound}) (fig~5).
\begin{figure}[t]
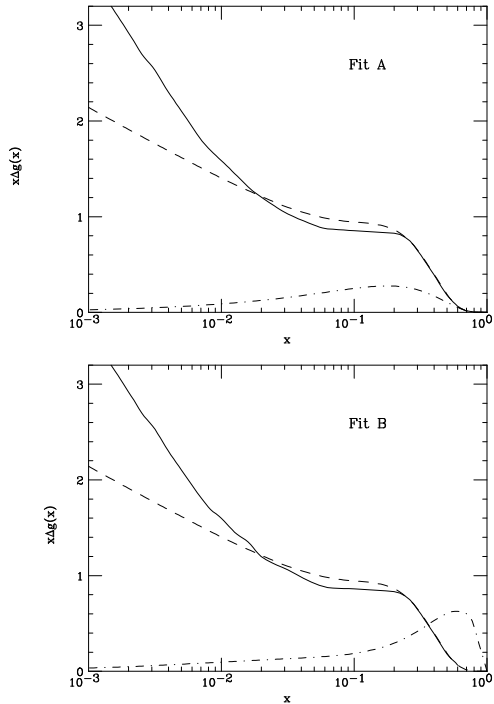

\vbox{
\hbox{\hskip.6truecm\epsfig{figure=fig3a.eps,width=6.truecm}}
\vskip.2truecm
\hbox{\hskip.6truecm\epsfig{figure=fig3b.eps,width=6.truecm}}}
\vskip-.6truecm
\caption{\footnotesize 
The maximal gluon distribution at $Q^2=1$~GeV$^2$ 
obtained from the LO (dashed lines) and
NLO  (solid lines) positivity
bounds using the polarized quark distributions corresponding to fits A
and B of ref.~[6]. The corresponding best-fit polarized gluon
distribution from ref.~[6] is also shown (dot-dashed).}\vskip-.8truecm
\end{figure}

Comparison of the NLO bounds on $\Delta g$ thus obtained from the
positivity constraint with the results of a recent NLO fit to $g_1$
data~\cite{ABFR} (fig.~3) underscores their phenomenological
relevance.
Indeed, the polarized gluon distribution can be currently
determined~\cite{ABFR}
only from scaling violations of $g_1$, 
\begin{eqnarray}&&
{d\over dt} g_1^{\rm singlet} (N,Q^2)=\\
&&\quad=
{\langle e^2\rangle\over2} {\alpha_s\over
2\pi} \left[\gamma_{qq}
\Delta \Sigma
+2n_f\gamma_{qg}\Delta g\right]+O(\alpha_s^2).\nonumber\end{eqnarray}
These, however, only allow a
determination of small moments of $\Delta g$, while high
moments are essentially unconstrained, due to the fact that at large $N$
$|\gamma_{qg}|\ll|\gamma_{qq}|$ (see fig.~4).
\begin{figure}[t]
\vbox{
\hskip.6truecm\epsfig{figure=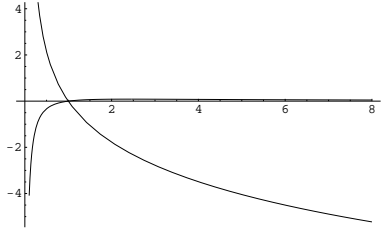,width=5.truecm}}
\vskip-.6truecm
\caption{\footnotesize 
The LO anomalous dimensions $\gamma_{qq}(N)$ and $\gamma_{qg}(N)$ (top to
bottom at small $N$) as a function of $N$.}\vskip-.8truecm
\end{figure}
As a consequence, the data can be described equally well by
polarized  gluon distributions whose higher moments are widely
different, provided their low moments remain more
or less the same. This is the case of
the $A$ and $B$ fits of ref.~\cite{ABFR} displayed in
fig.~3. The corresponding uncertainty is thus especially
noticeable in the large $x$ region. Whereas this is not important if
one is only interested in a determination of first moments, it
prevents a good determination of the shape of $\Delta g$. The
positivity bounds, by providing an independent constraint on higher
moments allow to substantially improve the determination of the shape
of $\Delta g$.

To this purpose it is important to understand at which scale the
bounds should be imposed. Indeed, it is easy to see that 
\begin{equation}\lim_{Q^2\to\infty}{|\Delta q_i(N,Q^2)|\over q_i(N,Q^2)}=
{|\Delta g(N,Q^2)|\over g(N,Q^2)}=0.
\end{equation}
It follows that the LO bounds eq.(\ref{naivebound}) 
will always be satisfied at large enough
scale, but conversely they will always be violated at a low enough
scale. Correspondingly, at low scale the NLO deviation from the LO
bounds will blow up. Hence, positivity bounds at large scales are
trivial, but they should not be imposed at
very low scale either, where they are unreliable.
However, if imposed at the boundary of validity of perturbation
theory, positivity bounds can be phenomenologically relevant in
providing complementary information which is useful in the
determination of the shape of polarized parton distribution.
\vglue-.4truecm
%BBBBBBBBBBBBBBBBBBBBBBBBBBBBBBBBBBBBBBBBBBBBBBBBBBBBBBBBBBBBBBBBBBBBBBBBBBB%

%\smallskip


\begin{thebibliography}{6}
\footnotesize
\bibitem{partsch} G.~Altarelli,
R.~K.~Ellis and 
G.~Martinelli, {\it  Nucl. Phys.} {\bf B143} (1978) 521; {\bf B157}
(1979) 461.
\bibitem{pos} G.~Altarelli, S.~Forte and G.~Ridolfi, {\tt hep-ph/9806345}
\bibitem{discf}
W.~A.~Bardeen et al, {\it Phys. Rev.} {\bf D18} (1978)
3998.\\
J.~Kodaira, {\it
Nucl. Phys.} {\bf B165} (1980) 129.
\bibitem{higgs} R.~K.~Ellis
et al., {\it Nucl. Phys.} {\bf B297} (1988) 221.\\
S.~Dawson, {\it Nucl. Phys} {\bf
B359} (1991) 283.\\
 A.~Djouadi, M.~Spira and P.~M.~Zerwas, {\it
Phys. Lett.} {\bf B264} (1991) 440.
\bibitem{cteq}H.~L.~Lai et
al., {\it Phys. Rev.}
{\bf D55} (1997) 1280.
\bibitem{ABFR} G.~Altarelli, R.~D.~Ball, S.~Forte and
G.~Ridolfi, \NP\vyp{B496}{1997}{337}; {\it Acta
Phys. Pol.} {\bf B29} (1998) 1145, 
{\tt hep-ph/9803237}.
\end{thebibliography}
\end{document}